\documentclass[12pt]{article}
\begin{document}
\def\scri{\unitlength=1.00mm
\thinlines
\begin{picture}(3.5,2.5)(3,3.8)
\put(4.9,5.12){\makebox(0,0)[cc]{$\cal J$}}
\bezier{20}(6.27,5.87)(3.93,4.60)(4.23,5.73)
\end{picture}}

\begin{center}

{\large THE HAWKING ENERGY ON PHOTON SURFACES}

\vspace{10mm}

{\large Ingemar Bengtsson}

\vspace{7mm}

{\sl Stockholms Universitet, AlbaNova\\
Fysikum\\
S-106 91 Stockholm, Sverige}

\vspace{5mm}

{\bf Abstract:}

\end{center}

{\small 

\noindent The Hawking energy has a monotonicity property under the inverse mean curvature 
flow on totally umbilic hypersurfaces with constant scalar curvature in Einstein spaces. 
It grows if the hypersurface is spacelike, and decreases if it is timelike. Timelike examples include Minkowski and de Sitter hyperboloids, and photon surfaces in Schwarzschild.}

\vspace{8mm}

{\bf 1. Introduction}

\

\noindent Without much fanfare, Hawking proposed a measure of the energy associated 
to a closed surface in spacetime \cite{Hawking}. We will explain it below, and 
refer elsewhere for more details \cite{Eardley, Szabados, Bray}. It is agreed 
that Hawking's expression does not have all the properties one expects `energy' 
to have, but it does have the single most important property of that concept: it 
is useful. Notably, an argument initiated by Geroch \cite{Geroch} and finalized 
by Huisken and Ilmanen \cite{Huisken} uses the Hawking energy of spheres embedded 
in a time-symmetric hypersurface to prove the Riemannian Penrose inequality. 
This is a great improvement in our understanding of energy in general relativity. 
The idea is to set up a geometric flow that moves any sphere to a large round sphere 
close to infinity, while all the time increasing the Hawking energy. From this 
point of view the occurence of negative Hawking energies in Minkowski space is not 
a drawback, it is a virtue. One can dream of a generalization of the Geroch flow 
to spacetime leading to a proof of the Penrose inequality in full generality 
\cite{Frauendiener}, but then one has to deal with the fact that Minkowski space 
also contains spheres with positive Hawking energy \cite{Mars}. Hence there must 
be a subtle story to tell about the flow and any monotonicity property 
that the Hawking energy has, if the dream is to come true. 

Our ambitions in this note are quite modest. First we show that the Hawking energy is 
never positive on spacelike hyperboloids in Minkowski space, and never negative on
timelike hyperboloids. (With a suitable definition the same holds for de Sitter 
space.) Then we show heuristically that, in Einstein spaces, the Hawking energy is 
monotone under Geroch's inverse mean curvature flow on any totally umbilic 
hypersurface with constant curvature. In particular it cannot increase if that 
hypersurface is timelike. Timelike totally umbilic hypersurfaces are known as 
photon surfaces, because they are swept out by light rays emitted tangentially 
from an embedded surface \cite{Claudel}. Spacetimes admitting complete photon 
surfaces are quite rare \cite{Perlick1}, but spherically symmetric spacetimes offer 
obvious examples (because then one can choose a round sphere to emit light rays 
from). There has been quite a bit of interest in photon surfaces recently 
\cite{Perlick2, Ong, Gibbons}. The papers by Cederbaum et al. are particularly 
relevant for us \cite{Carla1, Carla2, Carla3}. The photon surfaces that fill the 
Schwarzschild exterior have constant scalar curvature \cite{Carla1}, so that they 
provide examples to which our observation about the monotonicity of the Hawking 
energy applies. 

\vspace{6mm} 

{\bf 2. The Hawking energy}

\

\noindent Since we will deal with topological 2-spheres embedded in hypersurfaces that 
are themselves embedded in spacetime, there will be a bit of a strain on the 
notation. We will use $R_S, \bar{R}$, and $R$ for the scalar curvature of 
respectively the spheres, the hypersurfaces, and the spacetime, and we use $\gamma_{ab}, 
\bar{g}_{ab}$, and $g_{ab}$ for their respective first fundamental forms. The normal 
vector of a hypersurface is denoted by $\vec{e}$, and the normal vector of a surface 
within a hypersurface by $\vec{n}$. Timelike normal vectors are assumed to be future 
directed. The null normals of the surface are therefore 

\begin{equation} \vec{k}_\pm = \left\{ \begin{array}{lcc} \vec{e} \pm \vec{n} & & 
\mbox{if the hypersurface is spacelike} \\ \vec{n} \pm \vec{e} & & 
\mbox{if the hypersurface is timelike.} \end{array} \right. \end{equation}

\noindent The shape tensor of the surface is denoted by the kernel letter $K$, 
the second fundamental form of a hypersurface is denoted by the kernel 
letter $\Pi$, and that of the surface within the hypersurface by $p$. The trace 
of the shape tensor contracted by a normal vector is denoted with a subscript. Thus we 
will come across formulas such as $K_n = p$, meaning that the trace of the 
shape tensor contracted into the normal vector $\vec{n}$ is 
equal to the trace of the second fundamental form of the surface within 
the hypersurface. We do not think that the notation will cause any difficulties, 
but the fact that the hypersurface can be timelike or spacelike can be 
confusing. The shape tensor can be split into 

\begin{equation} K_{ab}(k_\pm ) = \sigma_{\pm ab}+ \frac{1}{2} \gamma_{ab} 
\theta_\pm \ , \end{equation}

\noindent where we introduced the null expansions $\theta_\pm$ and the traceless 
shears $\sigma_{\pm ab}$. 

The definition of the Hawking energy can now be stated in three equivalent forms: 

\begin{equation} E_H= \sqrt{\frac{A}{16\pi}}\left ( 1 + \frac{1}{16\pi}\oint \theta_+ 
\theta_- {\rm d}S  - \frac{\lambda}{3}\frac{A}{4\pi} \right) \ , \label{1} 
\end{equation}

\begin{equation} E_H= \frac{1}{16\pi} \sqrt{\frac{A}{16\pi}}\left( \oint 
( 2R_S + \theta_+ \theta_-) {\rm d}S  - \frac{\lambda}{3}\frac{A}{4\pi}\right) 
\ , \label{2} \end{equation} 

\begin{equation} E_H = \frac{1}{8\pi} \sqrt{\frac{A}{16\pi}}\oint \left( 
\sigma_{+ab}\sigma_-^{ab} +(G_{ab} + \lambda g_{ab})k_+^ak_-^b -
\frac{1}{2} C_{abcd}k_+^ak_-^bk_+^ck_-^d \right) {\rm d}S \ . 
\label{3} \end{equation}

\noindent Here $A$ is the area of the 2-sphere, $C_{abcd}$ is the Weyl tensor, 
and $G_{ab}$ the Einstein tensor. The term proportional to the cosmological 
constant $\lambda$ was added to Hawking's definition as an afterthought 
\cite{Boucher}. To go between (\ref{1}) and (\ref{2}), apply the Gauss--Bonnet 
theorem. To go between (\ref{2}) and (\ref{3}), use the Gauss formulas in the 
codimension 2 case. 

From (\ref{3}) we see immediately that the Hawking energy vanishes for an 
arbitrary cut of a lightcone in Minkowski and de Sitter space, because in 
these cases all the curvature terms and one of the shear tensors vanish. We 
can also make use of 

\begin{equation} \sigma_{+ab}\sigma_-^{ab} = \left\{ \begin{array}{lcc} 
\sigma_{eab}\sigma_e^{ab} - \sigma_{nab}\sigma_n^{ab} & & 
\mbox{if the hypersurface is spacelike} \\ \sigma_{nab}\sigma_n^{ab} 
- \sigma_{eab}\sigma_e^{ab} & & 
\mbox{if the hypersurface is timelike.} \end{array} \right. \end{equation}

\noindent A totally umbilic hypersurface is defined as one whose second 
fundamental form is everywhere shear-free. (The strange name `umbilic' is 
due to the fact that the surface of the human body has a shear-free second 
fundamental form at the centre of the navel.) On such a hypersurface it holds 
that 

\begin{equation} \sigma_{e ab} = 0 \ . \end{equation}

\noindent Inspection 
of formula (\ref{3}) then shows that the Hawking energy is never positive  
for arbitrary spheres in a totally umbilic spacelike hypersurface in a conformally 
flat Einstein space, and never negative if the hypersurface is timelike. On a 
timelike static cylinder both signs can occur \cite{Szabados, Mars}. For 
a round sphere in the Schwarzschild spacetime the Hawking energy evaluates to 
$m$, the total mass of the spacetime. 

\vspace{6mm} 

{\bf 3. The Geroch flow}

\

\noindent Geroch's idea was to move a sphere within a hypersurface, in the direction 
of its normal vector $\vec{n}$ and with a speed that depends on the extrinsic curvature 
of the sphere \cite{Geroch}. When doing so the metric and the mean curvature of the 
surface change according to 

\begin{equation} \dot{\gamma}_{ij} = {\cal L}_{\phi \vec{n}}\gamma_{ij} = 2 \phi 
p_{ij} \ \end{equation}

\begin{equation} \dot{p} = - \Delta_S\phi - \frac{1}{2}\phi (p_{ij}p^{ij} + p^2) 
+ \frac{\epsilon}{2}\phi (R_S - \bar{R}) \ . \end{equation}

\noindent The formula for $\dot{p}$ is the formula for the second variation of 
the area rewritten using the Gauss' equation, $\Delta_S$ is the intrinsic Laplacian, 
and $\epsilon = \vec{n}^2$. This gives us another sign to remember, 

\begin{equation} \epsilon = \left\{ \begin{array}{lcc} +1  & & 
\mbox{if the hypersurface is spacelike} \\ -1 & & 
\mbox{if the hypersurface is timelike.} \end{array} \right. \end{equation}
 
\noindent For the rate of the flow Geroch sets 

\begin{equation} \phi = \frac{1}{p} \ , \end{equation}

\noindent which is why his flow is referred to as the Inverse Mean Curvature Flow. 
With this choice $\dot{A} = A$. (To learn about curvature flows in general, consult 
Sethian \cite{Sethian}.) Geroch showed heuristically that the Hawking energy 
has an important monotonicity property if the flow takes place within a time 
symmetric hypersurface. If $p = 0$ at some point of the surface the flow can only 
exist in a suitable weak sense, which is why it took so long to turn Geroch's 
arguments into a rigourous theorem \cite{Huisken}. 

There have been many studies of the monotonicity properties of the Hawking energy 
in more general situations \cite{Eardley, Mars, BHMS}. Here we will apply the Geroch flow to a 
sphere within a totally umbilic hypersurface inside an Einstein space, but we allow 
the hypersurface to be timelike. Consider the Hawking energy in version (\ref{2}), that is 

\begin{equation} E_H = \frac{1}{16\pi}\sqrt{\frac{A}{16\pi}} \oint 
\left( 2 R_S + \epsilon K_e^2- \epsilon K_n^2 - \frac{4\lambda}{3} \right) {\rm d}S \ . \end{equation}

\noindent We make use of 

\begin{equation} K_e 
= \gamma^{ab}\Pi_{ab} = \frac{1}{3}\Pi \gamma^{ab}\bar{g}_{ab} = \frac{1}{3}\Pi 
(\bar{g}^{ab} - n^an^b)\bar{g}_{ab}= \frac{2}{3} \Pi \ . \end{equation}

\noindent Recall that $\bar{g}_{ab}$ and $\gamma_{ab}$ are the first fundamental 
forms of the hypersurface and the surface, respectively. We also use the equation 
$K_n = p$. The Gauss equation combined with the Einstein equation implies 

\begin{equation} \epsilon \bar{R} + \frac{2}{3}\Pi^2 = 2G_{ab}e^ae^b = 2\epsilon 
\lambda \ . \end{equation}

\noindent  Then the Hawking energy is 

\begin{equation} E_H = \frac{1}{16\pi}\sqrt{\frac{A}{16\pi}} \oint 
\left( 2 R_S - \epsilon p^2 - \frac{2}{3}\bar{R}\right) {\rm d}S \ . \end{equation}

\noindent We now repeat Geroch's calculation \cite{Geroch}. We only have an extra 
sign and an extra term involving $\bar{R}$ to keep track of. The result is 

\begin{equation} \dot{E}_H = \frac{1}{16\pi}\sqrt{\frac{A}{16\pi}} 
\left( \epsilon C - \frac{2}{3}\oint \dot{\bar{R}}{\rm d}S \right) \ , \end{equation}

\noindent where 

\begin{equation} C = \oint \left( \frac{2}{p^2}D_ipD^ip + 
\left(p_{ij}- \frac{p}{2}\gamma_{ij}\right) \left( p^{ij}-\frac{p}{2}\gamma^{ij}\right) 
\right) {\rm d}S \geq 0 \ . \end{equation}

\noindent If the scalar curvature $\bar{R}$ of the 
hypersurface is constant so that $\dot{\bar{R}} = 0$ then the Hawking energy 
is a monotone quantity. It can only increase 
if $\epsilon = +1$, and it can only decrease if $\epsilon = -1$. 

The derivation is heuristic because it leaves open the question whether 
the flow exists. On spacelike hypersurfaces this is a hard question \cite{Huisken}. 
Surfaces embedded in timelike hypersurfaces are in many ways less wild than those 
one finds embedded in spacelike hypersurfaces, but on a timelike hyperboloid in 
Minkowski space we do have the problem that through every point there passes a 
surface with $p = 0$. (The timelike hyperboloid is maximally symmetric, so it is 
enough to show this for one point.) But let us consider a round sphere with 
$p$ positive and constant over the surface, and let us normalize the hyperboloid 
in which it sits so that $\bar{R} = 6$. A quick calculation confirms that 
$p < 2$ for such a surface, and another quick calculation that 

\begin{equation} \dot{p} = \frac{2}{p} - \frac{p}{2} = 
\frac{(2-p)(2+p)}{2p} > 0 \ . \end{equation}

\noindent This evolves towards $p = 2$. Although this is a special case it seems clear 
that surfaces whose mean curvatures are everywhere positive will flow to round spheres 
at \scri \ without encountering any special problems. The problem with $p = 0$ 
will loom very large in the Schwarzschild photon sphere at $r = 3m$, but that is 
a very special case since the photon sphere never reaches \scri . 

\vspace{6mm}

{\bf 4. Photon surfaces}

\

\noindent All the photon surfaces in the Schwarzschild spacetime, not just the photon 
sphere, have constant scalar curvature \cite{Carla1}. This provides a reasonably large 
set of examples for which the Hawking mass decreases under the inverse mean curvature 
flow, and it is worthwhile to see exactly how it happens. 

We start with the metric 

\begin{equation} ds^2 = - V(r)dt^2 + \frac{dr^2}{V(r)} + r^2d\theta^2 + r^2\sin^2{\theta} 
d\phi^2 \ , \hspace{8mm} V(r) > 0 \ . \end{equation}

\noindent We define a spherically symmetric timelike hypersurface through 

\begin{equation} r = r(t) \ . \end{equation}

\noindent We denote its first fundamental form by $g_{ab}$, and define 

\begin{equation} F(r, \dot{r}) = V(r) - \frac{\dot{r}^2}{V(r)} > 0 \ . \end{equation} 

\noindent (Now the dot notation refers to differentiation with respect to $t$.) The 
second fundamental form is quickly computed to be 

\begin{equation} \Pi_{ab} = \frac{1}{\sqrt{F}}\frac{V}{r}g_{ab} + 
\frac{1}{\sqrt{F}}\left(\frac{V^2}{r} - \frac{VV_{,r}}{2} - \left( \frac{1}{r} - 
\frac{3}{2}\frac{V_{,r}}{V}\right) \dot{r}^2 - \ddot{r}\right)\nabla_at\nabla_bt \ . 
\end{equation}

\noindent The hypersurface is a photon surface if and only if the expression within 
brackets vanishes. This gives a second order differential equation for $r(t)$ which can 
be derived from a Lagranian. In principle it can be solved using quadratures, because 
it admits the conserved quantity 

\begin{equation} \frac{r^2}{V} - \frac{r^2\dot{r}^2}{V^3} = - 2E = r_0^2 \ . \end{equation}

\noindent The different cases that arise are discussed by Cederbaum 
and et al. \cite{Carla2, Carla3}. We can use the information we already have to compute 
the mean curvature 

\begin{equation} \Pi = g^{ab}\Pi_{ab} = \frac{3}{r_0} \end{equation}

\noindent and (somewhat more labouriously) the scalar curvature 

\begin{equation} \bar{R} = \frac{6}{r_0^2} 
+ \frac{2}{r^2}(1-V - V_{,r})  \end{equation}

\noindent of the photon surface. 

So far we have not imposed Einstein's equations. When we do we find that 

\begin{equation} V(r) = 1 - \frac{2m}{r} - \frac{\lambda}{3}r^2 \end{equation}

\noindent and we have arrived at the Schwarzschild-de Sitter spacetime. In this case 
we find that 

\begin{equation} \bar{R} = \frac{6}{r_0^2} + 2\lambda \ , 
\end{equation}

\noindent a constant value for all our photon surfaces. Indeed, in the Einstein case 
this follows directly from the fact that the mean curvature is constant. 

\vspace{6mm} 

{\bf 5. Envoi}

\

\noindent It is comforting that we have a set of examples to which our observation about the 
Hawking energy applies. Of course it has to be admitted that the Schwarzschild spacetime 
is a very special spacetime. The Penrose inequality holds in spherical symmetry \cite{Malec, 
Husain}, and we are not able to offer any suggestions about how to move beyond that. But 
we have seen that the Hawking energy has many subtle properties. Building intuition for 
them is likely to be useful. 

\vspace{10mm} 

\noindent \underline{Acknowledgements}: I thank Patrik Lindberg and Jos\'e 
Senovilla for sharing their many insights, and Carla Cederbaum for sketching the 
content of reference \cite{Carla3}. And I happily acknowledge the 
Mittag--Leffler Institute in Djursholm, Stockholm, for a wonderful 
relativity program.

\end{document}